\documentclass[a4paper,cite]{article}
\usepackage{latexsym}

\newcommand{\lbl}[1]{\label{eq:#1}}
\newcommand{ \rf}[1]{(\ref{eq:#1})}

\newcommand{\be}{\begin{equation}}
\newcommand{\ee}{\end{equation}}
\newcommand{\bea}{\begin{eqnarray}}
\newcommand{\eea}{\end{eqnarray}}
\newcommand{\setl}{\setlength\arraycolsep{2pt}}

\newcommand{\noi}{\noindent}
\newcommand{\nn}{\nonumber}
\newcommand{\ra}{\rightarrow}

\newcommand{\lesssim}{ {\
\lower-1.2pt\vbox{\hbox{\rlap{$<$}\lower5pt\vbox{\hbox{$\sim$}}}}\ } 
}
\newcommand{\gtrsim}{ {\
\lower-1.2pt\vbox{\hbox{\rlap{$>$}\lower5pt\vbox{\hbox{$\sim$}}}}\ } 
}

\newcommand{\cL}{{\cal L}}

\newcommand{\cO}{{\cal O}}

\newcommand{\Imm}{\mbox{\rm Im}}

\newcommand{\tr}{\mbox{\rm tr}}

\newcommand{\MeV}{\mbox{\rm MeV}}
\newcommand{\GeV}{\mbox{\rm GeV}}

\newcommand{\annd}{\mbox{\rm and}}

\newcommand{\hc}{\mbox{\rm h.c.}}

\newcommand{\al}{\alpha}
\newcommand{\als}{\alpha_{\mbox{\rm {\scriptsize s}}}}

\newcommand{\GF}{G_{\mbox{\rm {\tiny F}}}}

\newcommand{\gL}{\frac{1-\gamma_{5}}{2}}
\newcommand{\gR}{\frac{1+\gamma_{5}}{2}}

\newcommand{\h}{\mbox{\bf h}}

\newcommand{\stern}{\langle\bar{\psi}\psi\rangle}

\input epsf

\begin{document}

\begin{titlepage}

\begin{flushright} CPT-2001/P.4133\\ UAB-FT-508\\ \today
\end{flushright}
\vspace*{1.5cm}
\begin{center} 
{\Large \bf A Critical Reassessment of 
$Q_{7}$ and $Q_8$ Matrix Elements}\\[3.0cm]

{\bf Marc Knecht}$^a$, {\bf Santiago Peris}$^b$ and  {\bf Eduardo de
Rafael}$^a$\\[1cm]

$^a$  Centre  de Physique Th{\'e}orique\\
       CNRS-Luminy, Case 907\\
    F-13288 Marseille Cedex 9, France\\[0.5cm]
$^b$ Grup de F{\'\i}sica Te{\`o}rica and IFAE\\ Universitat Aut{\`o}noma
de Barcelona, 08193 Barcelona, Spain.\\

\end{center}

\vspace*{1.0cm}

\begin{abstract}

We compare recent theoretical determinations of weak matrix elements of
the electroweak penguin operators $Q_{7}$ and $Q_{8}$. We pay special
attention to the renormalization scheme dependence of these
determinations as well as to the influence of higher dimension operators
in the different approaches.

\end{abstract}

\end{titlepage}

\section{\normalsize Introduction}
\lbl{int}

\noi
There has been recent progress in understanding the
bosonization of some of the four--quark operators which appear in the
effective Lagrangian of the Standard Model describing
$K$ physics~\footnote{See refs.~\cite{KPdeR00} and~\cite{PdeR00} 
and references
therein.}. The bosonization of the electroweak penguin operators
\be
Q_7 = 6(\bar{s}_{L}\gamma^{\mu}d_{L})
\sum_{q=u,d,s} e_{q} (\bar{q}_{R}\gamma_{\mu}q_{R})\quad\annd\quad Q_8 =
-12\sum_{q=u,d,s}e_{q}(\bar{s}_{L}q_{R})(\bar{q}_{R}d_{L})
\ee
in particular, 
where $e_q$ denote quark charges in units of the electric charge and summation 
over quark colour indices within brackets is understood,
has been obtained
by two different analytic methods. One of the methods uses the framework of the
$1/N_c$ expansion~\cite{KPdeR99}; the other one combines dispersion 
relations with phenomenological input~\cite{DG00,N00}. Both methods are well
rooted in the underlying QCD theory, and 
therefore they are competitive with more
standard non--perturbative techniques based on lattice QCD simulations. In
fact, the existence of lattice QCD estimates of matrix
elements of the $Q_7$ and $Q_8$ operators~\cite{Detal99,C98,A98,LDL98} 
allows for a detailed comparison between these different approaches. It is
precisely this comparison which is the main concern of this letter. We shall
also discuss, within the particular case of matrix elements of the
$Q_7$ operator, the r\^{o}le of ``higher dimension operators'' in weak matrix
elements estimates which has been recently  raised in refs.~\cite{CDG00,D00}.
 
The operator
$Q_7$ emerges at
the
$M_W$ scale from  considering  the so--called electroweak penguin
diagrams. In the presence of the strong interactions, the renormalization 
group evolution of
$Q_7$ from the scale $M_W$ down to a scale $\mu\lesssim m_{c}$ mixes this
operator with the four--quark density--density operator $Q_8$, 
among others. The
$Q_8$ operator plays an important r\^{o}le in the phenomenology of
CP violation because, in the effective four--quark Lagrangian, it
appears modulated by a Wilson coefficient $C_{8}(\mu^2)$ which has a sizeable
imaginary part induced by the Kobayashi--Maskawa phase in the flavour mixing
matrix of the underlying Electroweak Model. 

It is well known~\cite{BW84} that the bosonization of the $Q_{7}$ and $Q_{8}$
operators leads to a term with no derivatives in the low energy effective chiral
Lagrangian. This Lagrangian generates $\vert\Delta S\vert\!=1\!$ transitions
among the pseudoscalar Goldstone fields of the spontaneously broken
$SU(3)_{L}\times  SU(3)_{R}$ symmetry of the QCD Lagrangian with three massless
flavours $u$, $d$,
$s$. It has the following form:
\be\lbl{order0}
\cL_{\vert\Delta S\vert=1}^{(0)}=-\frac{\GF}{\sqrt{2}}\frac{\al}{\pi}
V_{ud}V_{us}^{\ast}\,\frac{M_{\rho}^{6}}{16\pi^2}\,\h\,\tr
\left(U\lambda_{L}^{(23)}U^{\dagger}Q_{R}\right) +\hc \,,
\ee
where $U$ denotes the $3\times 3$ matrix field which collects the octet of
pseudoscalar Goldstone fields, $Q_{R}=\mbox{\rm diag}[2/3, -1/3, -1/3]$ is the
right--handed charge matrix associated with 
the electromagnetic couplings of the
light quarks and
$\lambda_{L}^{(23)}$ is the effective left--handed flavour matrix
$\left(\lambda_{L}^{(23)}\right)_{ij}=\delta_{i2}\delta_{3j}$ $(i,j=1,2,3)$.
Under chiral  rotations ($V_{L},V_{R}$):
\be
U\ra V_{R}UV_{L}^{\dagger}\,,\quad Q_{R}\ra V_{R}Q_{R}V_{R}^{\dagger}\,,\quad
\lambda_{L}^{(23)}\ra V_{L}\lambda_{L}^{(23)}V_{L}^{\dagger}\,,
\ee
and the trace on the r.h.s. of Eq.~\rf{order0} is an invariant.
Actually, this is the only possible invariant which can generate
$\vert\Delta
S\vert\!=\!1$ transitions in the Standard Model to order
$\cO(\GF\,\al)$ in the electroweak coupling and to $\cO(p^0)$ in the chiral
expansion. With the normalization chosen in Eq.~\rf{order0},
the coupling constant $\h$ is dimensionless. This constant plays a
major r\^{o}le in the phenomenological analysis of $K\ra\pi\pi$ amplitudes. It
is one of the basic couplings of the low energy effective electroweak
Lagrangian of the Standard Model that one would like to evaluate. 

The $1/N_c$ expansion in QCD~\cite{'tH74,W79} offers a specific
non--perturbative framework for discussing the 
dynamics which governs low energy
constants like $\h$ in Eq.~\rf{order0}. It was recently shown in
ref.~\cite{KPdeR99} that the contribution to the  constant
$\h$ from the $Q_7$ and $Q_8$ four--quark operators can be calculated
to first non--trivial order in the $1/N_c$ expansion.

At the theoretical level, 
the comparison between different approaches can be made 
by evaluating the same matrix elements of the operators $Q_7$
and $Q_8$ in each approach; provided of course that the calculations are made
at the same scale $\mu$ and within the same renormalization scheme. In this
respect it is important to remember, as
clearly emphasized in ref.~\cite{Buras98}, that it is {\it not} 
enough to just state 
the definition of $\gamma_5$ in $D$ dimensions, i.e. whether one uses 't
Hooft-Veltman (HV) or naive dimensional regularization (NDR) 
or any other definition. One
must also define the so-called evanescent operators and 
in this work we stick to the treatment and
convention of ref.~\cite{BuWei90}. In particular we emphasize that within
this convention Fierz symmetry {\it is} a valid concept even in $D$
dimensions.   

However, 
there is 
still a technical difficulty when comparing theoretical evaluations of matrix
elements. The calculations within the dispersive approach reported in
ref.~\cite{DG00}, although aimed at evaluating matrix elements at $\cO(p^0)$
in the chiral expansion, are not formulated within the large--$N_c$
framework; while the lattice evaluations cannot be restricted, at least at
present, to specific orders neither in the chiral expansion nor in the
$1/N_c$ expansion. Of course, if, as often claimed, the approximations
within each approach are {\it good}, then the results should agree with each
other within the
estimated theoretical errors.
Regretfully, the comparisons which have been made so far in the
literature correspond to estimates  which were obtained
in {\it different} schemes.  

As seen within the combined frameworks of the chiral expansion and the
$1/N_c$ expansion, the counting of possible contributions to 
matrix elements of
the
$Q_7$ and
$Q_8$ operators can be summarized as follows. Denoting by $\langle
Q_{7,8}\rangle\vert_{\cO(p^0)}$ the contribution to $\h$ in Eq.~\rf{order0}
from $Q_{7,8}$, one has
\be
\langle Q_{7}\rangle\vert_{\cO(p^0)}= \underline{\cO(N_c)}      +\cO(N_c^0)\,,
\qquad\annd\qquad
\langle Q_{8}\rangle\vert_{\cO(p^0)}= \underline{\cO(N_{c}^2)}+\cO(N_c^0)\,,
\ee
were only the underlined contributions were calculated in ref.~\cite{KPdeR99}.
The contribution of $\cO(N_c^0)$ to $\langle Q_{8}\rangle\vert_{\cO(p^0)}$ is
Zweig suppressed. As we shall see, it involves the sector of 
scalar (pseudoscalar)
Green's functions  where it is hinted from various 
phenomenological sources that
the restriction to just the 
leading large--$N_c$ contribution may not always be a good
approximation. It is precisely this issue which
prevented us in ref.~\cite{KPdeR99} from giving numerical values of the matrix
elements of the $Q_8$ operator. Here we shall follow a different
strategy that does not suffer from the above shortcomings, which will allow
us to make predictions for the matrix elements of the $Q_8$ operator as
well. 

\vspace{0.7cm}


\section{\normalsize Bosonization of $Q_{7}$ and $Q_{8}$}
\setcounter{equation}{0}
\lbl{bq7bq8}

\subsection{\normalsize\sc The $Q_{7}$ Operator}

\noi
Because of its left--right vector structure, the factorized  component of
the operator $Q_{7}$, which is $\cO(N_c^2)$, cannot contribute to order
$\cO(p^0)$ in the low--energy effective Lagrangian. The first
contribution from this operator of chiral $\cO(p^0)$  is at least of
$\cO(N_c)$ in the $1/N_c$ expansion and, formally, it can be written as
follows 
\be\lbl{LRgral}
Q_7 = 6\ \langle
O_{1}(\mu)\rangle
\,
\, 
\tr\left( U(x)\lambda_{L}^{(23)} U^{\dagger}(x) Q_{R}\right)^{\dag} \,,
\ee
where 
\be\lbl{Oone}
\langle O_1\rangle \equiv\langle 0\vert
(\bar{s}_{L}\gamma^{\mu}d_{L})(\bar{d}_{R}\gamma_{\mu}s_{R})\vert 0 
\rangle  \ .
\ee
Matrix elements of $Q_7$ are then given by the matrix elements of the
effective operator 
\be\lbl{chiraloperator}
\tr\left( U(x)\lambda_{L}^{(23)} U^{\dagger}(x)
Q_{R}\right)= \frac{2}{F_0^2} K^- \pi^+ + \frac{i \sqrt 2}{F_0^3} \left[\bar
K^0 \pi^- \pi^+ + \frac{1}{\sqrt 2} K^- \pi^+ \pi^0 + \dots \right]
\ee
on the r.h.s. of Eq.~\rf{LRgral} times the dynamical factor $6\langle O_1
\rangle$.  In particular, defining the isospin decomposition as ($a=7,8$) 
\bea\lbl{isospin}
\langle \pi^+\pi^- \vert Q_a\vert K^0\rangle&=& 
i \ \langle(\pi\pi)_{I=0}\vert Q_a\vert K^0\rangle + 
\frac{i}{\sqrt 2} \ \langle(\pi\pi)_{I=2}\vert Q_a\vert K^0\rangle \nn \\
\langle \pi^0\pi^0\vert Q_a\vert K^0\rangle&=& 
i \ \langle(\pi\pi)_{I=0}\vert Q_a\vert K^0\rangle - 
i \ \sqrt 2 \ \langle(\pi\pi)_{I=2}\vert Q_a\vert K^0\rangle \ , 
\eea
and taking into account that the operator of Eq.~\rf{chiraloperator} has
no transition between a $K^0$ and a two--$\pi^0$ state,   
we find that

{\setl
\bea
\langle(\pi\pi)_{I=2}\vert Q_{7}\vert K^{0}\rangle & = & 
\langle(\pi\pi)_{I=2}\vert Q_{7}^{(3/2)}\vert K^{0}\rangle=
\frac{-1}{F_{0}}\langle \pi^{+}\vert Q_{7}^{(3/2)}\vert
K^{+}\rangle=\frac{-1}{3F_{0}}\langle \pi^{+}\vert Q_{7}\vert K^{+}\rangle
\lbl{chiral7}
\\
 & = & 
- \frac{4}{F_0^3} \langle O_1\rangle \lbl{chiral7largen} \,, 
\eea}

\noi
where $Q_{7}^{(3/2)}$ denotes the isospin $I\!=\!3/2$ component of the
$Q_{7}$ operator, i.e.
\be
Q_{7}^{(3/2)}=2\bar{s}_{L}\gamma^{\mu}d_{L}
\left(\bar{u}_{R}\gamma_{\mu}u_{R}-\bar{d}\gamma_{\mu}d_{R}\right)+
2\bar{s}_{L}\gamma^{\mu}u_{L}\bar{u}_{R}\gamma_{\mu}d_{R}\,.
\ee
Notice that the relations in Eq.~\rf{chiral7} are exact symmetry properties of
the $\cO(p^0)$ term in the chiral expansion.

\subsection{\normalsize\sc The $Q_{8}$ Operator}

As pointed out in ref.~\cite{DG00}, independently of large--$N_c$
considerations, the bosonization of the
$Q_8$ operator to $\cO(p^0)$ in the chiral expansion 
can also be related to four--quark condensates by current
algebra Ward identities, with the result
\be\lbl{Q8leading}
Q_8 = - 12 \  \langle O_{2}(\mu)\rangle\,\,
\tr\left( U(x)\lambda_{L}^{(23)} U^{\dagger}(x) Q_{R}\right)^{\dag}\,,
\ee
where 
\be\lbl{Otwo}
\langle O_2 \rangle\equiv \langle 0\vert
(\bar{s}_{L}s_{R})(\bar{d}_{R}d_{L})\vert
0\rangle \, .
\ee
{F}rom this result, recalling Eqs.~\rf{isospin},
 there follows the corresponding matrix element relations

{\setl
\bea
\langle(\pi\pi)_{I=2}\vert Q_{8}\vert K^{0}\rangle & = & 
\langle(\pi\pi)_{I=2}\vert Q_{8}^{(3/2)}\vert K^{0}\rangle=
\frac{-1}{F_{0}}\langle \pi^{+}\vert Q_{8}^{(3/2)}\vert
K^{+}\rangle=\frac{-1}{3F_{0}}\langle \pi^{+}\vert Q_{8}\vert K^{+}\rangle
\lbl{chiral8}
\\
 & = & 
\frac{8}{F_{0}^3} \langle O_{2}\rangle\,,
\lbl{largenn}
\eea}

\noi 
where, again, the relations in Eq.~\rf{chiral8} are exact symmetry
properties of the
$\cO(p^0)$ term. The
operator $Q_{8}^{(3/2)}$ denotes the isospin $I\!=\!3/2$ component of the
$Q_{8}$ operator, i.e.
\be
Q_{8}^{(3/2)}=-4 (\bar{s}_{L}u_{R})(\bar{u}_{R}d_{L}) + 
4 (\bar{s}_{L}d_{R}) \left[ (\bar{d}_{R}d_{L})-(\bar{u}_{R}u_{L})\right]\,.
\ee

\vspace{0.7cm}

\section{\normalsize The large-$N_c$ approach}
\setcounter{equation}{0}
\lbl{largeN}

\subsection{\normalsize {\sc Evaluation of $\langle O_{1}\rangle$}}

As shown in ref.~\cite{KPdeR99} the vacuum expectation value $\langle
O_1\rangle$ can be expressed as an integral
\be\lbl{Ooneintegral}
\langle
O_1\rangle=\frac{1}{6}\left(-3ig_{\mu\nu} \int\frac {d^4q}{(2\pi)^4}
\Pi_{LR}^{\mu\nu}(q)\right)
\ee
involving the two--point function
\be\lbl{lrtpf}
\Pi_{LR}^{\mu\nu}(q)=2i\int d^4 x\,e^{iq\cdot x}\langle 0\vert
\mbox{\rm T}\left(L^{\mu}(x)R^{\nu}(0)^{\dagger}
\right)\vert 0\rangle\,,
\ee
with currents
\be L^{\mu}=\bar{q}_{i}(x)\gamma^{\mu}\gL\, q_{j}(x)
\quad \annd \quad
R^{\mu}=\bar{q}_{i}(x)\gamma^{\mu}\gR\, q_{j}(x)\,,
\ee
and  $i$ and $j$ fixed flavour indices, $i\not= j$. This is the same
two--point function which governs the electroweak $\pi^{+}-\pi^{0}$ mass
difference~\cite{KPdeR98}~; however, because of the factorization of the
short--distance contributions in the Wilson coefficient $C_{7}(\mu^2)$, the
integral that appears in Eq.~\rf{Ooneintegral} is divergent for
large $Q^2$.  Then, for consistency of the matching between the
long--distance evaluation which we are concerned with here, and the
short--distance evaluation, the integral in Eq.~\rf{Ooneintegral} should be
done in the same renormalization scheme as the calculation  of the Wilson
coefficients. The fact that we are interested in
$\cO(p^0)$ terms in the chiral Lagrangian implies furthermore that the
calculation must be done in the chiral limit. 
 
As a first step, we shall therefore use dimensional regularization with 
$D=4-\epsilon$, and define the integral in Eq.~\rf{Ooneintegral} as follows 
\be\lbl{LRgraldim}
\left(-3ig_{\mu\nu}\int \frac{d^4q}{(2\pi)^4}
\Pi_{LR}^{\mu\nu}(q)\right)_{D}\ra\frac{3(1-D)}{16\pi^2}\frac{(4\pi\mu^2)^{\,
\epsilon/2}} {\Gamma(2-\epsilon/2)}\int_{0}^{\infty}dQ^2
(Q^2)^{1-\epsilon/2}\left( -Q^2\Pi_{LR}(Q^2)\right)\vert_{D}\,,
\ee
where $Q^2= -q^2$ and 
\be\lbl{lritpf}
\Pi_{LR}^{\mu\nu}(q)=(q^{\mu}q^{\nu}-g^{\mu\nu}q^2)\Pi_{LR}(Q^2)\,.
\ee
The matching to short distances is controlled by the operator product
expansion (OPE)~\cite{SVZ79}, and in $D$ dimensions and in the large--$N_c$
limit one has 
\be\lbl{ope}
\lim_{Q^2\ra\infty} (1-D)Q^4\left(
-Q^2\Pi_{LR}(Q^2)\right)\vert_{D}=-12\pi^2\left(\frac{\als}{\pi}+
\cO(\als^2)\right)
\left[1+(\kappa-2/3)\frac{\epsilon}{2}\right]\stern^2\,,
\ee
where $\kappa$ depends on the renormalization scheme~\footnote{This is in
agreement with the calculation reported in ref.~\cite{DG00}.}: 
\be\lbl{kappa}
\kappa=-1/2\quad\mbox{\rm in NDR}\,,\quad\annd\quad\kappa=+3/2\quad\mbox{\rm
in HV}\,.
\ee

The next step is the use of the hadronic representation of
the spectral function associated with
$\Pi_{LR}(Q^2)$, and this is where the large--$N_c$ limit plays an
important simplifying r\^{o}le. In this limit the spectral 
function  consists of
the difference between
 an infinite number of narrow vector states and an infinite
number of narrow axial--vector states, together with the Goldstone pion pole:
\be
\frac{1}{\pi}\Imm\Pi_{LR}(t) =\sum_{V}f_{V}^2 M_{V}^2\delta(t-M_{V}^2)
-\sum_{A}f_{A}^2 M_{A}^2\delta(t-M_{A}^2)-F_{0}^2\delta(t)\,.
\ee
This spectral function is furthermore constrained by the fact that there are no
operators of dimension $d=2$ and $d=4$ which contribute to the OPE of
$\Pi_{LR}(Q^2)$, which in turn implies the two Weinberg sum
rules
\be\lbl{weinbergsrs}
\sum_{V}f_{V}^2 M_{V}^2-\sum_{A}f_{A}^2 M_{A}^2=F_{0}^2
\quad
\annd\quad
\sum_{V}f_{V}^2 M_{V}^4-\sum_{A}f_{A}^2 M_{A}^4=0\,.
\ee
{F}rom the fact that $\Pi_{LR}(Q^2)$ obeys an unsubtracted dispersion relation
it then follows that 
\be\lbl{hadrons}
-Q^2\Pi_{LR}(Q^2)=\sum_{A}\frac{f_{A}^2
M_{A}^6}{Q^2(Q^2+M_{A}^2)}-\sum_{V}\frac{f_{V}^2 M_{V}^6}{Q^2(Q^2+M_{V}^2)}\,.
\ee
Matching now the OPE of the function $\Pi_{LR}(Q^2)\vert_{D}$ in the
quark--gluon language, as given in Eqs.~\rf{ope} and \rf{kappa}, onto the
corresponding one in the hadronic language in Eq.~\rf{hadrons} requires
\be
\left(\sum_{V}f_{V}^2 M_{V}^6-\sum_{A}f_{A}^2
M_{A}^6\right)_{D=4-\epsilon}  =  
\left(\sum_{V}f_{V}^2 M_{V}^6-\sum_{A}f_{A}^2 M_{A}^6\right)_{D=4}
 \times \left(1+\kappa\frac{\epsilon}{2}\right)\,,
\ee
The integral in Eq.~\rf{LRgraldim} can now be made in the hadronic
representation of large--$N_c$ QCD. Using the $\overline{\mbox{\rm
MS}}$--renormalization scheme, we get the following result
\be\lbl{LRgralmsb}
\langle O_1\rangle=\left(-\frac{1}{2} ig_{\mu\nu}\int \frac{d^4q}{(2\pi)^4}
\Pi_{LR}^{\mu\nu}(q)\right)_{\overline{\mbox{\rm{\footnotesize
MS}}}}^{\mbox{\rm\tiny ren.}}= 
 -\frac{3}{32\pi^2}\left[\sum_{A}f_{A}^2
M_{A}^{6}\log\frac{\Lambda^2}{M_{A}^2}-\sum_{V}f_{V}^2
M_{V}^{6}\log\frac{\Lambda^2}{M_{V}^2}
\right]\,,
\ee
where 
\be\lbl{cutoff}
\Lambda^2=\mu^2\exp(1/3+\kappa)\,,
\ee
and $\kappa$ has been defined in Eq.~\rf{kappa}.
Notice that in ref.~\cite{KPdeR99} another scheme was used, 
different from NDR and HV, in which $\kappa=0$.

\subsection{\normalsize {\sc Evaluation of $\langle O_{2}\rangle$}}

In the $1/N_c$ expansion one has that 
\be\lbl{blind}
\langle O_2 \rangle= \frac{1}{4}
\left[\langle \bar{\psi}\psi(\mu)\rangle^2 +\cO(N_c^0)\right]\,,
\ee
and therefore $\langle O_2 \rangle$ is directly related to the quark
condensate
$\stern$, provided we restrict ourselves to the first term in this
expression which is $\cO(N_c^2)$.
However, as already mentioned in the introduction, there are reasons to
suspect that subleading terms in the $1/N_c$ expansion involving some
Green's functions with scalar (and pseudoscalar) density operators might be
important, unlike those which only involve vector (and axial-vector)
currents~\footnote{There are several phenomenological examples of this: the
$\eta'$ mass, the possible existence of a broad $\sigma$ meson, large final
state interactions in states with $J=0$ and $I=0$,
etc...~\cite{Moussallam00}  In order to study
this issue in a systematic way we are considering at
present the possibility that the
appropriate expansion for these exceptional Green's functions could be a 
$1/N_c$ expansion in which $n_{f}/N_c$ is held fixed, where $n_f$ denotes 
the number of
light flavours. This kind of  expansion was originally 
advocated by G.~Veneziano~\cite{V76}.}. Consequently 
one should be
cautious about neglecting the
$\cO(N_c^0)$ contributions in Eq.~\rf{blind} and consider instead 
\be
\langle O_{2}\rangle=\frac{1}{4}\stern^2+
\langle(\bar{s}_{L}s_{R})(\bar{d}_{R}d_{L})
\rangle_{\mbox{\rm c}}\,.
\ee
The unfactorized contribution involves Feynman diagrams which require
gluon exchanges between at least two quark loops. These are 
the so called Zweig--suppressed contributions, which are indeed
$\cO(N_c^0)$ in the $1/N_c$ expansion. This ``subleading'' term is
governed by a two--point function of the type ($Q^2=-q^2\,,$ ,   
and $i,j=u,d,s$ with $i\ne j$)
\be
\Psi_{ij}(Q^2)=i\int d^4 x e^{iq.x}\langle 0\vert
T\left\{\bar{q}_{i}(x)\gR q_{i}(x)\,\,
\bar{q}_{j}(0)\gL q_{j}(0)\right\}\vert 0\rangle\,.
\ee
More precisely
\be\lbl{zweig}
\langle(\bar{s}_{L}s_{R})(\bar{d}_{R}d_{L})
\rangle_{\mbox{\rm c}}=\frac{1}{i}\left(\int\frac{d^D
q}{(2\pi)^D}\,
\Psi_{ds}(Q^2)\right)_{\overline{\mbox{\rm{\footnotesize
MS}}}}^{\mbox{\rm\tiny ren.}}\,,
\ee
where the integral should again be defined in the same 
renormalization scheme
as the short--distance calculation of the Wilson coefficients; 
which explains the
meaning of the $\overline{\rm MS}$ subscript. There is, in particular, a
contribution to this integral from the singlet
$\eta^{(0)}$ pseudoscalar, which in the chiral limit acquires 
a mass because of
the axial $U(1)$ anomaly. This contribution, which very likely 
plays an important r\^ole in the 
low--$Q^2$ regime of the integral in 
Eq.~\rf{zweig}, can easily be calculated with the result
\be
\langle(\bar{s}_{L}s_{R})(\bar{d}_{R}d_{L})
\rangle_{\eta^{(0)}}=  - \frac{M_{\eta^{(0)}}^2}{16\pi^2 F_{0}^2}\,
\frac{\stern^2}{6} \left(\log
\frac{\mu^2}{M^2_{\eta^{(0)}}}+1\right)\,,
\ee
in agreement with results reported in ref.~\cite{Ham00},
indicating the existence of $\cO(N_c^0)$ terms which could be potentially 
important. In this respect 
we wish to point out that the failure to properly
incorporate $\cO(N_c^0)$ Zweig--suppressed 
contributions is also a problem which
may seriously  affect the {\it quenched} lattice calculations of 
matrix elements of $Q_{8}$.

To allow for the possibility of large deviations
from the naive factorization in Eq.~\rf{blind}, it has become conventional
to assume an ansatz of the type~\cite{BNP92} 
\be\lbl{fudge}
\langle O_{2}\rangle (\mu) = \ \frac{\rho(\mu)}{4} \ \stern^2(\mu)\ ,
\ee
where $\rho(\mu)$ parametrizes
possible deviations from the leading $\cO(N_c^2)$ factorization where
$\rho \to 1$. The crucial observation~\cite{DG00} is that, on the one hand, 
the same vev $\langle
O_{2}\rangle$ also appears in the OPE which
governs the high--$Q^2$
behaviour of the $\Pi_{LR}(Q^2)$ function (at $D=4$ and to lowest
order in the perturbative evaluation of the Wilson coefficient~\cite{SVZ79}), 
\be\lbl{ope2}
\lim_{Q^2\ra\infty} \left(
-Q^2\Pi_{LR}(Q^2)\right)Q^4  =  4\pi^2\frac{\als}{\pi}\left(4\langle
O_{2}\rangle+\frac{2}{N_c}\langle O_{1}\rangle\right)+\cO
\left(\frac{\als}{\pi}\right)^2\,.
\ee
On the other hand, we know
that large--$N_c$ QCD gives a reliable description for the $\Pi_{LR}(Q^2)$
function since it involves the vector and axial--vector channels only. In
fact, we can use the
renormalization group to resum this lowest--order result into~\cite{KPdeR99}
\be\lbl{Piresum}
\lim_{Q^2\to \infty}\left(-Q^6 \Pi_{LR}(Q^2)\right)=
\frac{4 \pi^2}{-\beta_1}\ \rho(Q) \ \widehat{\stern}^2
 \left(\frac{-\beta_1 \alpha_s(Q)}{\pi}\right)^
{\frac{2 \gamma_1+\beta_1}{\beta_1}} \ ,
\ee
where we have neglected the term proportional to $\langle O_{1}\rangle$, an
approximation which we shall justify {\it a posteriori}. Here,
$\widehat{\stern}=
\stern(\mu)  (\log\mu/\Lambda_{\overline {MS}})^{-9/22}$
denotes the scale invariant quark condensate and $\rho(Q)$ also includes the
effect of next--to--leading perturbative corrections.   Inserting the
large--$N_c$ expression which follows from Eq.~\rf{hadrons} in the l.h.s.
of Eq.~\rf{Piresum} results (at large $Q^2$) in 
\be\lbl{Pireson}
\sum_V f_V^2 M_V^6 - \sum_A f_A^2 M_A^6 \simeq   
\frac{4 \pi^2}{\beta_1}\ \rho(Q) \ \widehat{\stern}^2
 \left(\frac{-\beta_1 \alpha_s(Q)}{\pi}\right)^
{\frac{2 \gamma_1+\beta_1}{\beta_1}} \,;
\ee
and therefore
\be\lbl{otwo}
\langle O_2\rangle(\mu) \simeq  
\frac{1}{16
  \pi \alpha_s(\mu)}\left(\sum_A f_A^2 M_A^6 - \sum_V f_V^2 M_V^6 \right) \ . 
\ee
It is worth noticing the remarkable $Q^2$ independence of the r.h.s. of
Eq.~\rf{Pireson} due to the near cancellation of the exponent $\frac{2
 \gamma_1+\beta_1}{\beta_1}=2/11$ and the fact that $\rho(Q)\to 1$ as $Q^2\to
\infty$. Since the l.h.s. is clearly $Q$ independent, this is obviously a
welcome result.

We would like to pause here and comment on the $\cO(\alpha_s^2)$ term in
Eq.~\rf{ope2}. It is our understanding that  
the only existing calculation in the literature of this term
~\cite{Chetyrkin86} was done in a {\it  different} scheme of  evanescent
operators from the one of ref.~\cite{BuWei90}. Since this $\cO(\alpha_s^2)$
term is sensitive to the precise definition of evanescent
operators~\cite{Chetyrkin94}
\footnote{The analysis of ref.~\cite{Chetyrkin94} shows that this
 term of $\cO(\alpha_s^2)$ can actually be reduced in half by a redefinition
 of the operator basis in the $4-D$ extra dimensions.} and since
matrix elements of $Q_{7,8}$ eventually will have to be used in conjunction
with Wilson coefficients to obtain a physical result, it is of the utmost
importance that the extraction of $\langle O_2\rangle$ in Eq.~\rf{ope2} be
done within the same convention for evanescent operators as that employed in
the calculation of Wilson
coefficients. This is why we think, unlike the authors of  
refs.~\cite{DG00,N00}, that it is misleading to use the existing
calculation of the $\cO(\alpha_s^2)$ 
term in Eq.~\rf{ope2} and, consequently, we refrain from including it 
to obtain $\langle O_2\rangle$ in
Eq.~\rf{otwo}. This fact immediately introduces a major difference with
respect to refs.~\cite{DG00,N00} as the contribution of $\cO(\alpha_s^2)$
they take turns out to be  large, namely $\sim 50 \%$.

Numerical estimates for
 $\langle O_{1}\rangle$ in Eq.~\rf{LRgralmsb} and $\langle
O_{2}\rangle$ in Eq.~\rf{otwo} will be  made in the last section 
using the {\it minimal hadronic
ansatz} approximation to large--$N_c$ QCD (MHA). This approximation consists in
limiting the large--$N_c$ spectrum of narrow states to the minimal number
required to satisfy the OPE constraints relevant to the process that one is
considering. In our case, this requires the consideration of one vector state
and one axial--vector state, besides the 
pion pole~\footnote{ Sometimes we have 
referred to this particular case as the {\it
lowest meson dominance} approximation (LMD)~\cite{PPdeR98}.}. 
The MHA approximation has
recently been shown~\cite{PPdeR01} to successfully reproduce sum rules
which use the LEP experimental data of the ALEPH collaboration on hadronic
$\tau$ decay. This leads to a contribution of  $\langle
O_1\rangle$ to the r.h.s. of Eq.~\rf{ope2} 
which is indeed much smaller than the
one obtained from $\langle O_2\rangle$ in either 
scheme NDR or HV, justifying a
posteriori the initial approximation  
which was made to obtain Eq.~\rf{otwo}. 

\vspace{0.7cm}

\section{\normalsize The dispersive approach and 
the r\^{o}le of ``higher dimension operators''}
\setcounter{equation}{0}
\lbl{dispersive}

When computing $\langle O_1\rangle$ through Eq.~\rf{Ooneintegral}, 
the authors of ref.~\cite{DG00} have chosen to split the integral in
the regions $0\le Q^2\le \mu^2$ and $\mu^2 \le Q^2 \le \infty$ where $\mu$
plays the role of a sharp momentum cutoff, 
i.e. not to be confused with the scale
appearing in dimensional regularization. The low--$Q^2$ part 
is amenable to the use of the experimental data through a dispersion
relation. The high--$Q^2$ part, on the other hand, is divergent and must be
renormalized, e.g., in dimensional regularization in $D=4-\epsilon$ dimensions
with minimal subtraction (${\overline{MS}}$).
In this way one obtains
\be\lbl{dono1}
\langle O_1\rangle^{\overline {MS}}(\mu)=
\langle O_1\rangle^{\rm cutoff}(\mu)+ 
\frac{(1-D)}{32\pi^2}\frac{(4\pi\mu^2)^{\,
\epsilon/2}} {\Gamma(2-\epsilon/2)}\int_{\mu^2}^{\infty}dQ^2
(Q^2)^{1-\epsilon/2}\left( -Q^2\Pi_{LR}(Q^2)\right)\ ,
\ee
where $\langle O_1\rangle^{\rm cutoff}(\mu)$ corresponds to the low--$Q^2$
part, i.e.
\be\lbl{dono2}
\langle O_1\rangle^{\rm cutoff}(\mu)= \frac{3}{32\pi^2}\int_{0}^{\infty} 
dt\, t^2 \log\frac{t+\mu^2}{t}\frac{1}{\pi}\Imm\Pi_{LR}(t)\ .
\ee
Notice that, when writing Eq.~\rf{dono2}, one is using the first and
second Weinberg sum rules.

We are now in the position to discuss the issue of ``higher dimension
operators'' in the calculation of weak matrix elements which has been recently
raised in refs.~\cite{CDG00,D00}. On the one hand, these authors have
stressed the fact that the long--distance evaluation of weak matrix
elements should be done in the same renormalization scheme as the
short--distance evaluation of the Wilson coefficients. Although this is a
well known fact, they observe, quite rightly, that most of the available
long--distance calculations of weak matrix elements are done in a
cut--off scheme while using the $\overline{\rm MS}$ results of the Wilson
coefficients. Exceptions to that are the lattice QCD simulations and the
large--$N_c$ analytic results in refs.~\cite{KPdeR00,PdeR00}. On the
other hand, a more subtle issue also raised in refs.~\cite{CDG00,D00} is
the fact that in doing integrals of Green's functions which govern the
long--distance evaluation of weak matrix elements, in the same
$\overline{\rm MS}$ renormalization as used for the Wilson coefficients, one is
still confronted with the fact that, in principle,  the full string of terms in
the OPE contribute to these integrals because the integration over the euclidean
momentum goes all the way to infinity. We entirely agree with this observation 
but we want to show how the large--$N_c$ approach we have been advocating
avoids this criticism while the dispersive approach, as used in
ref.~\cite{DG00}, fails to incorporate the effect of higher 
dimension operators.

We consider first the evaluation 
of the vev $\langle O_{1}\rangle^{\overline{MS}}$ in
Eq.~\rf{dono1} to illustrate the point. Depending
on the input for $\Pi_{LR}$ in the integral over the high--$Q^2$ region in this
equation, one may obtain a unsatisfactory determination of $\langle
O_{1}\rangle^{\overline{MS}}$ even if $\langle O_{1}\rangle^{\rm cutoff}$ 
is derived reliably from the experimental data. The authors of
ref.~\cite{DG00} input for $\Pi_{LR}$ its {\it leading} asymptotic behaviour
coming from dimension six operators in the OPE, 
i.e. the equivalent to our large--$N_c$ Eq.~\rf{ope}. While
for $\mu$ large enough this is certainly a reasonable thing to do, for
the realistic value of $\mu=2$ GeV the
question still remains as to how large the contribution from higher dimension
operators in the OPE may actually get to be~\cite{CDG00,D00}.

In this regard we would like to point out that, in the
large--$N_c$ approach, the integrand in Eq.~\rf{hadrons}, if expanded in
powers of $1/Q^2$ gives not only the leading power of the OPE, which we use
as a constraint, but also the higher powers. As a matter of fact, and this is
the crucial point,  
in ref.~\cite{PPdeR01} it is shown that the 
MHA approximation to large--$N_c$ 
discussed at the end of section~\rf{largeN} does a
remarkably good job in predicting these higher vev's as compared to the ALEPH
data. Therefore one can use the MHA approximation to 
confidently evaluate the size of the contribution from 
higher dimension operators to the high--$Q^2$ region in the integral of
Eq.~\rf{dono1}. One then obtains

{\setl
\bea
\langle O_{1}\rangle^{\overline{MS}}(\mu)&=&\langle O_{1}\rangle^{\rm
  cutoff}(\mu)- \frac{3}{2\pi} \alpha_s \langle O_2\rangle
\left(\frac{1}{3}+\kappa\right) \lbl{master1}\\
&&-\frac{3}{32\pi^2}\left(\sum_V f_V^2 M_V^6 \log \frac{\mu^2+M_V^2}{\mu^2}-
  \sum_A f_A^2 M_A^6 \log \frac{\mu^2+M_A^2}{\mu^2}\right) \ ,\lbl{master2}
\eea}

\noi
where the expression \rf{master2} reads in the MHA approximation
($g_A=M_V^2/M_A^2$)  
\be\lbl{mhamaster2}
\simeq -\frac{3}{32\pi^2} \frac{F_0^2 M_V^4}{1-g_A} \left(\log
  \frac{\mu^2+M_V^2}{\mu^2} - 
\frac{1}{g_A} \log \frac{\mu^2+\frac{M_V^2}{g_A}}{\mu^2}\right) \ .
\ee
Equation \rf{master1} is the result of ref.~\cite{DG00} (in our notation),
with only the contribution from dimension--six operators. The 
expression in Eq.~\rf{mhamaster2}, which represents 
the contribution from operators of dimension 8 and higher, indeed goes to zero 
as an inverse power for large values of $\mu$, as expected. 
However it amounts to a $\sim 50 \%$ reduction at
$\mu=2$ GeV in NDR (see next section). 
We emphasize that in ref.~\cite{PPdeR01}
the MHA approximation was shown to yield correct predictions in the OPE 
up to operators of dimension 10, so we consider Eq.~\rf{mhamaster2} to be
reliable at least up to this order. 

The corresponding 
discussion in the case of $\langle O_2\rangle$ starts at Eq.~\rf{ope2}
with the contribution from $\langle O_1\rangle$ neglected, as already
explained. Therefore $\langle O_2\rangle$  is the operator modulating the
$1/Q^6$ fall-off in the OPE of the function $\Pi_{LR}$. At large $N_c$ one
obtains
\be \lbl{largeNcotwo}
\langle O_2\rangle \approx \frac{C(Q^2)}{16 \pi \alpha_s(\mu)} 
\left[ -Q^6 \Pi_{LR}(Q^2)\right] \ ,
\ee
where

{\setl
\bea\lbl{c}
C(Q^2)&=&\left[1- \frac{\sum_A \frac{f_A^2 M_A^8}{Q^2+M_A^2}-\sum_V \frac{f_V^2
      M_V^8}{Q^2+M_V^2}}{\sum_A f_A^2 M_A^6-\sum_V f_V^2 M_V^6} \right]^{-1} 
\nn \\
&\simeq&
\left[1- \frac{M_V^2}{Q^2+M_V^2} \frac{Q^2(g_A+1)+M_V^2}{g_A Q^2+M_V^2} 
\right]^{-1} \ ,
\eea}

\noi
where the last expression is a consequence of the MHA approximation. Again 
$C(Q^2)-1$ vanishes as an inverse power at high $Q^2$. Since in
refs.~\cite{DG00,N00} the contribution from higher dimension operators is
neglected, these authors effectively 
take $C(Q^2)=1$. However a typical value for $C(Q^2)$ is $\sim 40\%$ above
unity for $Q= 2\,\GeV$ (see next section for details).

\vspace{0.7cm}

\section{\normalsize Numerical Estimates and Comparisons}
\setcounter{equation}{0}
\lbl{ncc}

\noi
Using  Eqs.~\rf{LRgralmsb}, \rf{cutoff}, \rf{kappa}
 and \rf{otwo} one obtains in the
MHA approximation to large--$N_c$ QCD the simple expressions 

{\setl
\bea
\langle O_1\rangle (\mu) &=& -\ \frac{3}{32 \pi^2} \ \frac{F_0^2 M_V^4}{1-g_A}
\ \log\left[g_A 
\left(\frac{g_A \Lambda^2}{M_V^2}\right)^{\frac{1}{g_A}-1}\right] 
\lbl{oonemha}\\
\langle O_2\rangle (\mu) &=& \frac{1}{16 \pi \alpha_s(\mu)} 
\ \frac{F_0^2 M_V^4}{g_A} \, \, . \lbl{otwomha}
\eea}

\noi
For their numerical evaluation we shall 
take~\cite{PPdeR01}
\be
M_{V}=(750\pm 25)\,\MeV\,,\quad 
F_{0}=(87\pm 3)\,\MeV\,,\quad g_A= 0.50 \pm 0.06\,,
\ee 
and
$\als(\mu\!=\!2\,\GeV)=0.33\pm 0.04\,.$
Then the previous expressions lead to 
 
{\setl
\bea
\langle O_1\rangle_{(\mu=2\,\mbox{\rm
\tiny GeV})}  &=& \left.\begin{array}{cc} 
(-1.9\pm 0.2)\times 10^{-5}\ \GeV^6 & \mbox{\rm in
NDR}\\ (-1.1\pm 0.2)\times 10^{-4}\ \GeV^6 & \mbox{\rm in ~~HV}
\end{array}\right\}\lbl{oonenha}\\
\langle O_2\rangle_{(\mu=2\,\mbox{\rm
\tiny GeV})} &=& (2.9 \pm 0.6) \times 10^{-4}\ \GeV^6 
\,\, \mbox{\rm both in NDR and HV} \,\, ,\lbl{otwonha}
\eea}

\noi
which  in turn imply, through Eqs.~\rf{chiral7}, \rf{chiral7largen}, 
\rf{chiral8} and \rf{largenn}, 
\be
\langle(\pi\pi)_{I=2}\vert Q_{7}\vert K^{0}\rangle_{(\mu=2\,\mbox{\rm
\tiny GeV})}=\left.\begin{array}{cc} (0.11\pm 0.01)\,\GeV^3 & \mbox{\rm in
NDR}\\ (0.67\pm 0.09)\,\GeV^3 & \mbox{\rm in ~~HV}
\end{array}
\right\}\,,
\ee
and
\be
\langle(\pi\pi)_{I=2}\vert Q_{8}\vert K^{0}\rangle_{(\mu=2\,\mbox{\rm
\tiny GeV})}=(3.5\pm 0.8)\,\GeV^3\, .
\ee

In order to facilitate the direct comparison with the lattice results 
we also give the equivalent result for the matrix elements 
\be
\langle \pi^{+}\vert Q_{7}^{(3/2)}\vert K^{+}\rangle_{(\mu=2\,\mbox{\rm
\tiny GeV})}=\left.\begin{array}{cc} (-9.8\pm 0.6)\times 10^{-3}\,\GeV^4 &
\mbox{\rm in NDR}\\ (-5.8\pm 0.8)\times 10^{-2}\,\GeV^4 & \mbox{\rm in ~~HV}
\end{array}
\right\}\, ,
\ee
and 
\be\lbl{resultsMQ8}
\langle \pi^{+}\vert Q_{8}^{(3/2)}\vert K^{+}\rangle_{(\mu=2\,\mbox{\rm
\tiny GeV})}= (-0.30\pm 0.07)\,\GeV^4\,\, .
\ee
These are our predictions for the $O_{1,2}$ condensates and for the 
$Q_{7,8}$ matrix elements. The errors quoted come only from the propagation 
 from the input values for $M_V$, $F_0$, $g_A$ and $\alpha_s(2\,\GeV)$. 
We remark that, unlike the case of $Q_7$, matrix elements of
$Q_8$ do not show any dependence on the renormalization scheme (NDR vs. HV) at 
this level. 

In  
Table 1 we give a joint comparison of our results with the existing different
evaluations of matrix elements with which we can compare scheme dependences
explicitly~\footnote{This explains why, in this table, 
we only quote the lattice results of ref.~\cite{Detal99}. Model 
dependent calculations of the so called
$B$ factors associated with the $Q_{7,8}$ operators can also be
found,  in order of increasing
sophistication, in
refs.~\cite{Ber00},~\cite{Ham00} and~\cite{Pr00}.}. 

\begin{table*}[h]
\caption[Results]{Summary of matrix elements 
$M_{7,8}\equiv \langle(\pi\pi)_{I=2}\vert Q_{7,8}\vert K^{0}\rangle_{
(2\,{\mbox{\rm\tiny GeV}})} $ using naive dimensional regularization
(NDR)  and the 't Hooft-Veltman scheme (HV), in units of GeV$^3$.}
\lbl{table1}
\begin{center}
\begin{tabular}{|c|cccc|}
\hline
\hline 
Matrix Elements  & $M_7$(NDR) & $M_7$(HV) & $M_8$(NDR) &
$M_8$(HV)\\
\hline refs.~\cite{DG00,N00} & $0.22\pm 0.05$    &    &  $1.3\pm 0.3$  &
\\
\hline ref.~\cite{Detal99}& $0.11\pm 0.04$ & $0.18\pm 0.06$ &$ 0.51\pm
0.10$ & 
$0.62\pm 0.12 $\\ \hline 
This work (see also ref.~\cite{KPdeR99}) & $0.11\pm 0.03$ & $0.67\pm 0.20$
& 
$3.5\pm 1.1$ & $3.5\pm 1.1$ \\ \hline
\hline
\end{tabular}
\end{center}
\end{table*}

\noi 
In the table 
we have rounded off the errors of our predictions 
to an overall $30\%$, which we believe to be
a generous estimate of the systematic errors in our approach~\cite{GP01}.
For the case of $M_8$ this has the caveat that it of course assumes that the
$\cO(\alpha_s^2)$ corrections to Eq.~\rf{ope2}, once computed in the right
operator basis~\cite{BuWei90}, will be of a reasonable size.

We find
that, within the combined errors, our results for $M_{7}$ are in agreement
with the lattice results in the NDR scheme, but not in the HV
scheme \footnote{Although there is a difference bewteen what the Rome and
Munich groups call the "HV" scheme \cite{Buras98}, we have checked that the 
values for $M_{7}(HV)$ quoted in
the table are indeed consistent with the ones obtained from
$M_{7}(NDR)$ and $M_{8}(NDR)$ and the mixing of $Q_{7}$ and $Q_{8}$
under the appropriate change of scheme;
therefore, the disagreement with the lattice result for $M_{7}(HV)$ is
correlated to the strong discrepancy we have with the lattice result
for $M_{8}(NDR)$.}, 
while
they disagree by a factor of two with the dispersive results. 
The origin of this
disagreement can in fact be traced back to the effect of higher terms in the 
OPE found in Eq.~\rf{mhamaster2} which, as
discussed in the previous section, have been ignored in the
dispersive approach. As to the results of 
$M_{8}$, we are in disagreement with both the lattice and the dispersive
results. The discrepancy with the former may originate in the fact that most
of the contribution comes from an OZI--violating Green's function which is 
something inaccessible in the quenched approximation. As a matter of fact, 
lattice results are compatible with a value $\rho\sim 1$ in Eq.~\rf{fudge}
whereas we are finding that $\rho\sim 6$ for $\stern (2\,\GeV) \sim 
 (- 0.240\,\GeV)^3$, which is also the phenomenological result found in
ref.~\cite{DHGS98}. On the other hand, we do agree with the lattice results
on the independence of scheme (NDR vs. HV) in
$M_8$. As to the discrepancy on
$M_8$  with the dispersive results, it can be traced back 
to the $\sim 50\%$ $\cO(\alpha_s^2)$
correction they use for the Wilson coefficient in Eq.~\rf{ope2} 
plus the $\sim 40 \%$ effect coming
from the contribution from higher dimension operators in Eq.~\rf{c}. 


\indent

\noindent
{\large{\bf Acknowledgments}}

\vspace{0.3cm}

We thank Kostja Chetyrkin for informative communications on the work of
refs.~\cite{Chetyrkin86, Chetyrkin94}, Vicente Gimenez on the work of
ref.~\cite{Detal99}, Thomas Hambye
on ref.~\cite{Ham00} and Laurent Lellouch on ref.~\cite{LDL98}. We also thank
Hans Bijnens and Ximo Prades for discussions. This work
has been supported in part by TMR, EC-Contract No. ERBFMRX-CT980169
(EURODA$\phi$NE). The work of S.~Peris has also been partially supported by
the research project CICYT-AEN99-0766.



\end{document}